\documentclass[prb,twocolumn, amsmath,amssymb,showpacs,
 aps,groupedaddress]{revtex4}

\usepackage{hyperref}
\usepackage{color}
\usepackage{textcomp}
\usepackage{amsthm}
\usepackage{helvet}
\usepackage[T1]{fontenc}
\usepackage{times}
\usepackage{bm}
\usepackage{CJK}
\usepackage{eso-pic}
\usepackage{wasysym}
\usepackage{subfigure}
\usepackage{graphicx}
\usepackage{dcolumn}
\usepackage{bm}

\begin{document}

\title{Pressure Tuning of Collapse of Helimagnetic Structure in Au$_2$Mn}

\author{I-Lin Liu$^{1, 2, 3}$}
\email{ilin610@umd.edu}
\author{Maria J. Pascale$^{2}$}
\author {Juscelino B. Leao$^{1}$}
\author {Craig M. Brown$^{1}$}
\author {William D. Ratcliff$^{1}$}
\author {Qingzhen Huang$^{1}$}
\author{Nicholas, P. Butch$^{1,2}$}

\affiliation{
$^{1}$NIST Center for Neutron Research, National Institute of Standards and Technology, Gaithersburg, MD, 20899 USA\\
$^{2}$Center for Nanophysics and Advanced Materials, Department of Physics,  University of Maryland, College Park, MD 20742 USA\\
$^{3}$Department of  Materials Science and Engineering, University of Maryland, College Park, MD 20742-2115 USA}

\date{\today}
\begin{abstract}
We identify the phase boundary between spiral spin and ferromagnetic phases in Au$_2$Mn at a critical pressure of 16.4 kbar, as determined by neutron diffraction, magnetization and magnetoresistance measurements. The temperature-dependent critical field at a given pressure is accompanied by a peak in magnetoresistance and a step in magnetization. The critical field decreases with increasing temperature and pressure. The critical pressure separating the spiral phase and ferromagnetism coincides with the disappearance of the
magnetroresistance peak, where the critical field goes to zero. The notable absence of an anomalous Hall effect in the the ferromagnetic phase is attributable to the high conductivity of this material.

\end{abstract}

\pacs{Valid PACS appear here}
\maketitle


\section{Introduction}
Helimagnets have become popular materials in which to study spin textures such as skyrmions as well as antiskyrmions and tuning of spin-orbit coupling and the Dzyaloshinskii-Moriya interaction (DMI) \cite{Dzyaloshinskii1958,Moriya1960} typically as a change in a critical field ($H_c$), chemical composition and pressure. Topological properties such as stability of spin texture and  electronic control of skyrmions by varying bias voltage and current inspire researchers to further design novel applications\cite{Naoto2013, Wataru2015}. The helimagnetic phase diagrams in temperature and chemical composition share a similar pattern, with the spiral spin/ helix phase at low field, distorted spiral magnetic structure in intermediate field, and ferromagnetic phase at higher field above some critical field\cite{X. Z. Yu2010, Naoto2013}. The spin-orbit interaction plays an important role in the origin of novel magnetic states and phenomena such as giant magnetoresistance, topological insulators, and skyrmions. Modulated magnetic models have been discussed as the origin of helimagnets; the related modulated magnetic structures include ferromagnetic exchange energy, DMI, and quadruple spin-spin coupling, which affect the spin angle $\phi$ of the in-plane net spin between neighboring layers\cite{U1961, Nagamiya1962}.

    Au$_2$Mn has been long studied for its magnetic spiral spin structure\cite{A1956}, which recent calculations suggest is stabilized by electron correlations\cite{Glasbrenner2014}. As shown in Fig.~\ref{SA}, reports from neutron diffraction disagree regarding whether the temperature-dependent spin angle exhibits a minimum of $47^{\circ}$ at 120 K or the propagation vector maintains a constant value with spin angle $45^{\circ}$ below 120 $K$\cite{Handstein2005}. The giant magnetoresistance greater than $7\%$ and metamagnetic transition \cite{Hiroaki1998} showed that this is a tunable magnetic material with strong conduction electron interactions. Based on the pressure dependence of the magnetization, the critical pressure of the spiral spin to ferromagnetism (SS-FM) transition was predicted to be above 20 kbar\cite{Wayne1969}. Identifying the SS-FM phase boundary as a function of temperature and pressure provides insight into the variations of spin texture and competition between ferromagnetic exchange interaction and spin-orbit coupling.

    We present measurements of the temperature- and pressure-dependence of the helical spin angle, and the pressure dependence of the critical field of the SS-FM transition through magnetization and magnetoresistance (MR) measurements. Pressure drives Au$_{2}$Mn through a second order phase transition, which agrees with recent band structure calculations using DFT within the local spin density approximation (LSDA)\cite{Glasbrenner2014, Glasbrenner2016}. An anomalous magnetic peak in positive MR at low temperature and step-like magnetization track the SS-FM transition, and interestingly, in the pressure-induced FM phase, we did not see evidence of the expected Anomalous Hall Effect (AHE)\cite{NKanazawa2016, Toshiaki2015}.

\section{Experimental Methods}
Au$_{2}$Mn polycrystalline samples were prepared from high purity (6N) starting materials by arc melting in an argon atmosphere. Samples were flipped several times to ensure homogeneity and were further annealed for four days at $690\,^{\circ}{\rm C}$ under argon to yield samples with higher phase purity. The magnetization curve was taken by SQUID magnetometer from 2 K to 300 K up to 14 T. The magnetic susceptibility measured with a SQUID magnetometer confirmed the Neel temperature of 363 K\cite{HERPINA1961}.

    For the study of nuclear and magnetic structures, neutron diffraction experiments were performed on Au$_{2}$Mn at the NIST Center for Neutron Research (NCNR, Gaithersburg, USA), on the high resolution powder diffractometer BT1 with Ge311 as a monochromator with wavelength 2.079 {\AA}. The polycrystalline samples were placed in a HW-03 pressure cell with maximum pressure 10 kbar. To achieve the best hydrostaticity, a helium pressure transmitting medium was used. 

    For magnetotransport (MR) measurements, we placed a 1.3 $mm$ $\times$ 2.4 $mm$ square bar and 15 $\mu m$ thick sample of Au$_{2}$Mn curing contacts with silver epoxy into a nonmagnetic piston-cylinder pressure cell, using a 1 : 1 ratio of n-Pentane to 1-methyl-3-butanol as the pressure medium. Measurements taken from 300 to 1.8 K and in magnetic fields up to 8 T were performed below 16.6 kbar at 2 K (20.4 kbar at 300 K) in a commercial cryostat.

    Error bars in this paper correspond to an uncertainty of one standard deviation.
\section{Results and Discussion}

\begin{figure}
\centering
\subfigure{\label{SA}\includegraphics[scale=0.6, trim= 38 260 380 40,clip]{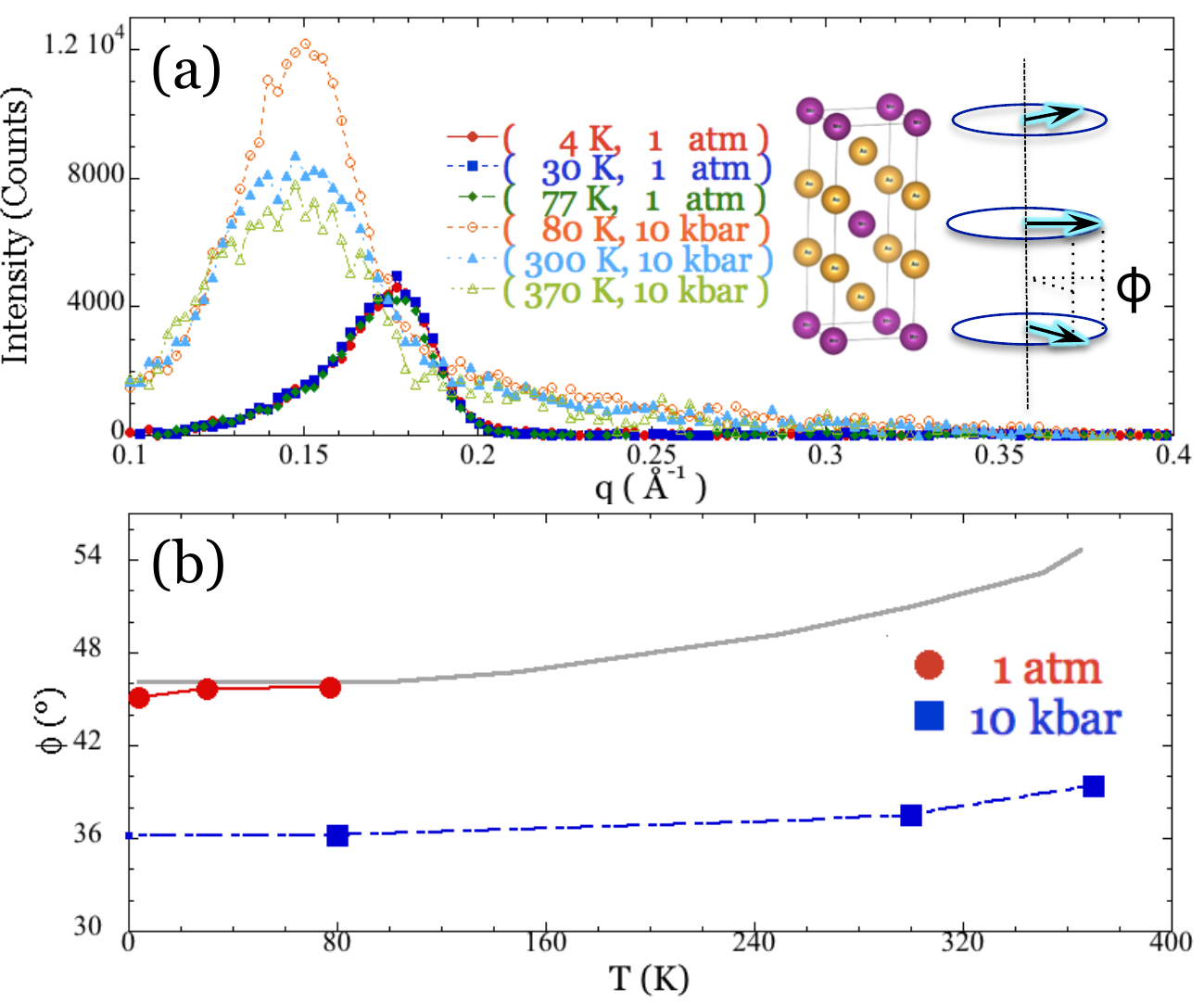}}
\subfigure{\label{SA-b}}
  \caption{( a ) The scattering vector of magnetic $000^{+}$ peak moves from 0.18 {\AA}$^{-1}$ to 0.15 {\AA}$^{-1}$ and the spin angle shrinks to $38^{\circ}$ when pressure rises to 10 kbar. The insert shows the tetragonal lattice structure and the corresponding spiral spin magnetic structure, which the in-plane spin locates at Mn atoms. ( b ) A comparison of the temperature dependence of spin angle determined from the $(0,0,0)^+$ peak at ambient pressure (red dot)  and 10 kbar (blue square). The gray line follows data from Ref\cite{Handstein2005}. The uncertainties are smaller than the marker size.}
\end{figure}
\begin{figure}
\centering
\subfigure{\label{MH}\includegraphics[scale=.43,trim= 35 40 180 35,clip]{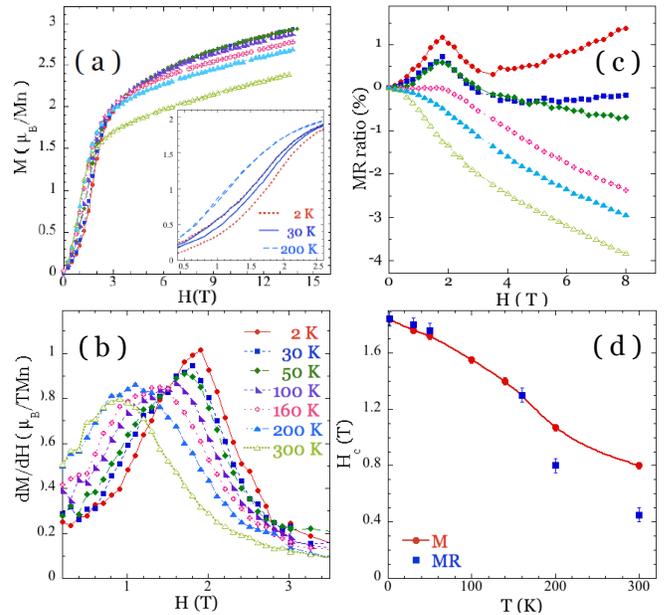}}
\subfigure{\label{MR-1atm}}
\subfigure{\label{dMdH}}
\subfigure{\label{HcT}}
  \caption{ ( a ) Magnetization and ( c ) the first derivative of magnetization of Au$_{2}$Mn up to 14 T. The inset highlights the distorted spiral transition zone between 0.2 T and 2.6 T. The hysteresis loop becomes narrow as temperature increases. ( b ) The MR ratio of Au$_{2}$Mn at ambient pressure. Below 50 K, the peak around 1.8 T indicates the SS-FM phase transition. This critical field of the SS-FM transition in Au$_{2}$Mn decreases as temperature increases. ( d ) The Temperature dependence of the critical field obtained from magnetization (red line) and MR (blue square) at ambient pressure. The agreement below 150 K suggests that the MR is a good probe of the low-temperature SS-FM phase boundary under pressure.}
  \label{Au2Mn-MH}
\end{figure}
\begin{figure}
\centering
\subfigure{\label{Au2Mn-13}\includegraphics[scale=.44,trim= 37 40 180 32,clip]{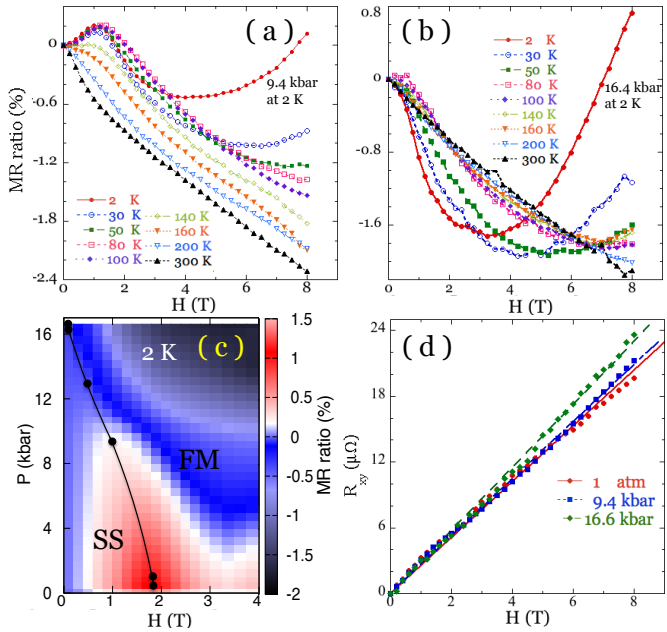}}
\subfigure{\label{Au2Mn-20}}
\subfigure{\label{Au2Mn-d2}}
\subfigure{\label{Hall}}
  \caption{Temperature dependence of MR ratio in Au$_{2}$Mn under ( a ) 13.2 kbar and ( b ) 20.5 kbar (at room temperature). ( c ) Pressure dependence of MR of Au$_2$Mn at 2 K. The black dotted line is the critical field defined by the field-induced peak in the MR, which indicates the SS-FM phase transition as a function of pressure and field. The red region has relatively large positive MR near the magnetic transition at 1.8 $T$ at ambient pressure, and the positive MR peak at low field is gradually suppressed by a critical pressure of 16.4 kbar. ( d ) Hall resistance of Au$_{2}$Mn at 2 K. The critical field decreases with pressure and is suppressed above the 16.4 kbar, critical pressure.}
\label{Au2Mn-MR}
\end{figure}

\begin{figure}
\centering
\includegraphics[scale=1.1,trim= 90 215 510 230,clip]{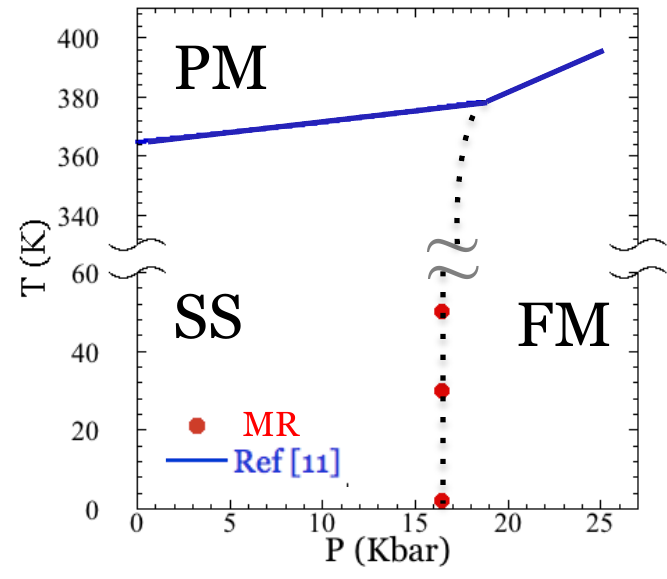}
  \caption{P-T phase diagram of Au$_{2}$Mn showing the SS-FM phase boundary (red dots) determined through MR measurements, Neel temperature and Curie temperatures(blue line)\cite{GRAZHDANKINA1963}. The almost vertical  SS-FM phase boundary is demarcated by a  dashed line.}
\label{PD}
\end{figure}

    We first address the pressure dependence of the magnetic spiral, which gives rise to magnetic satellite neutron diffraction peaks. As shown in Fig.~\ref{SA}, the scattering vector of the magnetic $000^{+}$ peak moves from 0.18 {\AA}$^{-1}$ to 0.15 {\AA}$^{-1}$ while pressure rises from ambient pressure to 10 kbar, which means the period of the spiral structure extends from 7 $nm$ to 8.3 $nm$. Figure~\ref{SA-b} shows the comparison of the corresponding spin angle at ambient pressure (red dot) and 10 kbar (blue square), for which $0.1^{\circ}$ error bars are within the markers. This result agrees with reference~\cite{Handstein2005, Inga2016} at 1 $atm$ that reports a spin angle $45.9^{\circ}$ as temperature falls  below 100 K (gray line in Fig.~\ref{SA-b}), which means that the spiral structure is not sensitive to temperature in this range. The consistency of the spin angle confirms that the magnetic properties of our sample prepared by argon arc melting are consistent with samples prepared by melt-spinning.

    The magnetic $000^{+}$ peak shifts to lower momentum transfer $q$, and $\phi$ contracts from $45.8^{\circ}$ to $38^{\circ}$ while pressure rises from ambient pressure to 10 kbar. Between 80 K and 370 K, $\phi$ at 10 kbar only increases by $3.2^{\circ}$, compared with the $10.4^{\circ}$ increase over the same temperature range at ambient pressure, which shows that high pressure enhances the rigidity of the magnetic structure. The existence of the magnetic $000^{+}$ peak, confirms that Au$_{2}$Mn keeps a spiral magnetic structure under 10 kbar below 370 K. 
  
    The temperature dependence of the magnetization at ambient pressure shows the transition from SS to FM phase as shown in Fig.~\ref{MH}.  The magnetic structure is spiral spin at zero field, turns into a distorted spiral while the applied field is less than the critical field and finally polarizes to ferromagnetism above the critical filed. The insert figure zooms in on the transition zone emphasizing a narrow magnetic hysteresis. The hysteresis loop shrinks by fifty percent when temperature increases from 2 K to 30 K. The magnetic hysteresis vanishes when temperature warms above 200 K. In figure~\ref{dMdH}, the first derivative of magnetization shows that the transition peaks move to weaker field as temperature rises from 2 K to 300 K. 

    At ambient pressure, the Hamiltonian of Au$_{2}$Mn is dominated by its exchange energy proportional to $\cos \phi$, and the critical field marks where the spiral structure collapses and forms ferromagnetism. Therefore, the corresponding critical field is roughly proportional to $\cos \phi$\cite{U1961, Nagamiya1962}, which explains why the spin angle increases as the critical field decreases with increasing temperature.  

    Figure~\ref{MR-1atm} shows the MR ratio ($MR=\frac{\rho(T)-\rho(0)}{\rho(0)} $) of Au$_{2}$Mn measured at ambient pressure. The positive MR ratio at 2 K (red circle), 30 K (blue square) and 50 K (green diamond) show the magnetic peak at 1.8 T indicating the SS-FM transition. The anomalous magnetic peak in the field dependence of the MR is related to the phase transition from SS-FM and its corresponding critical field. The low field positive MR in the SS phase transitions to negative MR in the spin-polarized FM phase, yielding a peak at the critical field observed in magnetization. The critical field is defined by the local maximum of the low field MR, which corresponds to a peak at low temperature, and the end of a plateau at high temperature. The transition peak becomes a plateau above 160 K, and the MR ratio becomes more negative with increasing  temperature. Similar low-temperature MR peaks have been observed on the border of the helical state in elemental holmium\cite{Akhava2014}. 
    
    The temperature dependence of the critical field has similar trends up to 200 K as measured by magnetization (red circle) and MR (blue square) which agree with each other below 160 K as shown in figure~\ref{HcT}. Although magnetic hysteresis was found in the magnetization, we do not observe hysteretic behavior in the MR. Above 160 K, the correspondence between the MR and magnetization becomes worse, possibly because the spin scattering is complicated by thermal excitations.

    Tracking the pressure dependence of the anomalous peak in MR and its related critical field is conducive to mapping the pressure dependence of the transition between SS-FM. The temperature dependence of MR in Au$_{2}$Mn under 9.4 kbar in the SS phase is presented in figure~\ref{Au2Mn-13}. The critical field at 2 K decreases to 1.2 T and the area below peaks shrinks. In figure~\ref{Au2Mn-20}, when pressure is increased above 16.4 kbar at 2 K, the magnetic peak is totally suppressed and MR is negative at low field. A positive monotonically increasing higher-field MR is observed below 50 K at all pressures. This may be associated with saturation of the magnetization.

    Both the critical field and magnetic structure are temperature-independent below 50 K. We display contour mapping of the MR at 2 K in the P-H plane in figure~\ref{Au2Mn-d2}. The black dotted line is the critical field defined by the field-induced peak in MR, which indicates the phase transition between SS-FM as a function of pressure and field. The red region has relatively large positive MR near the magnetic transition at 1.8 $T$ at ambient pressure, and the transition zone represented as the surrounding pink region is compressed as pressure increases. When pressure is close to 16.4 kbar, the transition peak is suppressed and gradually changes to negative magnetoresistance.
    
    The absence of an anomalous Hall effect is notable. The linear normal Hall resistance dominates the Hall resistance under pressure at 2 K as shown in figure~\ref{Hall}. The carrier density decreases smoothly from $6.144\times 10^{28} m^{-3}$ to $5.447\times 10^{28} m^{-3}$ when pressure rises from 1 $atm$ to 16.6 kbar. Although Au$_{2}$Mn becomes ferromagnetic above a critical pressure of 16.4 kbar, we do not see any qualitative changes in the Hall resistance as a function of pressure. In particular, we do not observe the expected AHE from itinerant ferromagnetism under pressure\cite{Nagaosa2010}. 

    The high conductivity of Au$_{2}$Mn appears to be the main reason why we do not detect the AHE. The Hall conductivity in the high-purity regime with high conductivity $\sigma_{xx}>$ 5 $\times 10^5( \Omega cm )^{-1}$\cite{Nagaosa2010}, is very challenging to investigate experimentally. In a high conductivity region with relative long mean free path ($l$), the ordinary Hall effect contribution dominates $ \sigma_{xy}$ and is proportional to $l^2$, while skew scattering which leads to AHE is proportional to $l$. Indeed, the  conductivity $\sigma_{xx}$ in our sample is larger than $10^6( \Omega cm )^{-1}$ at low temperature, which is at least one order higher than the high-purity limit. The large carrier density and conductivity minimize the scattering leading to the AHE. This high conductivity may also be related to the smaller GMR ratio in our sample relative to literature values\cite{Hiroaki1998}, which can be ascribed to larger grain size in our annealed samples\cite{Inga2016}.

    Figure~\ref{PD} shows the SS-FM transition as red dots in the pressure versus temperature phase diagram of Au$_{2}$Mn. These points correspond to the pressure at which the critical field is completely suppressed at low temperature (Fig.~\ref{Au2Mn-20}). The blue line shows Curie and Neel temperature\cite{GRAZHDANKINA1963}, having a kink around 18 kbar. The black dashed line shows the phase boundary between SS-FM, which is almost a temperature-independent, vertical line. This seems to be tied to the c-axis lattice constant. The thermal expansion coefficient for the c-axis at 10 kbar is only half the value at ambient pressure, which means that lattice constant of c-axis is $8.8617$ {\AA}  increasing with $8.3\times 10^{-5}$ {\AA} $K^{-1}$ from 80 K to 300 K. Meanwhile, the compressibility at 80 K is $6.23\times 10^{-3}$ {\AA}/ kbar from 10 kbar to ambient pressure, which means that the lattice is much more responsive to pressure. This agreement indicates that the MR is sensitive to the phase boundary under high pressure and low temperature. 

    Whether or not the SS-FM phase boundary is continuous is an interesting question. The nearly vertical SS-FM phase boundary suggests that the transition occurs discontinuously as a function of pressure. On the other hand, the decay of the MR peak with increasing pressure and smooth suppression of the critical field suggest that the transition is continuous. These observations support the notion that competing exchange terms in Au$_{2}$Mn put the SS-FM transition on the boundary between first- and second-order character\cite{Glasbrenner2014}. Considering the dependence of $\phi$ on competing exchange terms\cite{Glasbrenner2016}, for angles smaller than 30 degrees, the range of parameter space is limited, and even if $\phi$ smoothly goes to zero, it may occur over a narrow pressure range. Our results show that this question should be addressed by a neutron diffraction experiment using a pressure cell capable of exceeding 10 kbar.
    
\section{Conclusion}
We presented the SS-FM phase boundary determined via MR. The corresponding critical pressure is 16.4 kbar below 50 K. Neutron diffraction confirms that the spin angle decreases from $45^{\circ}$ to $36.18^{\circ}$ while pressure increases to 10 kbar. The critical field decreases as temperature rises at ambient pressure. The field-induced peak in MR is a great indicator of the SS-FM transition as a function of pressure. The critical field decreases, and transition zone also simultaneously shrinks as pressure rises. Below 50 K, the critical field is 1.8 T at ambient pressure and drops to 0 T at 16.4 kbar. 

\section{Acknowledgements}
The authors would like to acknowledge Yasuyuki Nakajima and Christopher J.  Eckberg for their help and discussion of pressure transport work. We thank J. K. Glasbrenner for discussion of neutron diffraction. Maria J. Pascale was sponsored by the Center for High Resolution Neutron Scattering as part of the NIST Summer Research Fellowship program, No. NSF DMR 1508249.

\end{document}